\documentclass[sigconf]{acmart}

\usepackage{tabularx}
\usepackage{multirow}
\newcolumntype{Y}{>{\centering\arraybackslash}X}
\AtBeginDocument{}
\begin{document}

\title{SceneTTS-Bench: A Benchmark for Scene-Level TTS \\ in Drama Dubbing}
\titlenote{This work was completed with support from Beijing DeepLogic
Intelligence Technology Co., Ltd. in March 2026 and submitted to the ACM
Multimedia (ACM MM) 2026 Dataset Track.}

\author{Yizhong Geng}
\authornote{Yizhong Geng and Yanliang Li contributed equally to this work.}
\affiliation{%
  \institution{Beijing University of Posts and Telecommunications}
  \city{Beijing}
  \country{China}}
\email{yzgeng@bupt.edu.cn}

\author{Yanliang Li}
\authornotemark[2]
\affiliation{%
  \institution{Beijing DeepLogic Intelligence Technology Co., Ltd.}
  \city{Beijing}
  \country{China}}
\email{liyanliang@luoji.cn}

\author{Jinghan Yang}
\affiliation{%
  \institution{Beijing University of Posts and Telecommunications}
  \city{Beijing}
  \country{China}}
\email{jinghanyang@bupt.edu.cn}

\author{Tianhan Jiang}
\affiliation{%
  \institution{University of California}
  \country{USA}}
\email{tij007@ucsd.edu}

\author{Yingming Gao}
\affiliation{%
  \institution{Beijing University of Posts and Telecommunications}
  \city{Beijing}
  \country{China}}
\email{yingming.gao@bupt.edu.cn}

\author{Ya Li}
\authornote{Corresponding author.}
\affiliation{%
  \institution{Beijing University of Posts and Telecommunications}
  \city{Beijing}
  \country{China}}
\email{yli01@bupt.edu.cn}

\renewcommand{\shortauthors}{Geng et al.}



\begin{abstract}
Text-to-speech systems are increasingly used for drama dubbing, yet
evaluation protocols remain sentence-level, leaving critical scene-level
behaviors insufficiently measured. We present SceneTTS-Bench, a benchmark
that evaluates TTS along three dimensions: timbre consistency across
character turns, emotional expressiveness on high-tension utterances, and
rhythm coherence under segmented long-form synthesis. The corpus comprises
real-world and generated drama scripts totaling 160 bilingual scenes
($\sim$10{,}300 utterances), with real-world scripts serving as the
primary source (100 scenes) and generated scripts as a supplementary
source (60 scenes), demonstrating the framework's extensibility through
synthetic data augmentation. A backend-agnostic Canonical Intermediate
Representation (Canonical IR) ensures fair cross-system comparison. Three
automatic pipelines produce per-utterance diagnostics: Speaker Consistency
Score (SCS) for timbre-drift detection, Under-Acting Ratio (UAR) for
under-acting identification, and Rate Discontinuity Ratio (RDR) for
rate-discontinuity quantification. Experiments on four TTS systems confirm
that each system exhibits distinct weaknesses, and that scene-level
rankings diverge substantially from sentence-level metrics. Benchmark
resources are publicly available at
\url{https://piedpiperg.github.io/scenetts-bench/}.
\end{abstract}

\maketitle

\section{Introduction}

Drama dubbing has become a widespread production need on streaming and
social-media platforms, including formats such as short-form drama, and TTS
systems are increasingly called upon to perform this task \cite{federico-etal-2020-speech}. In this setting
the role of TTS shifts from reading text to performing
characters: a system must hold a stable voice for each role across many
dialogue turns, deliver emotional intensity on dramatically pivotal
utterances, and avoid rhythm discontinuities when long utterances are split
for synthesis. These three demands, timbre consistency, emotional
expressiveness, and rhythm coherence, are scene-level properties that
surface only when utterances are judged jointly within a multi-role
dialogue, and they call for evaluation at a matching granularity.

\begin{figure*}[htbp]
  \centering
  \includegraphics[width=\textwidth]{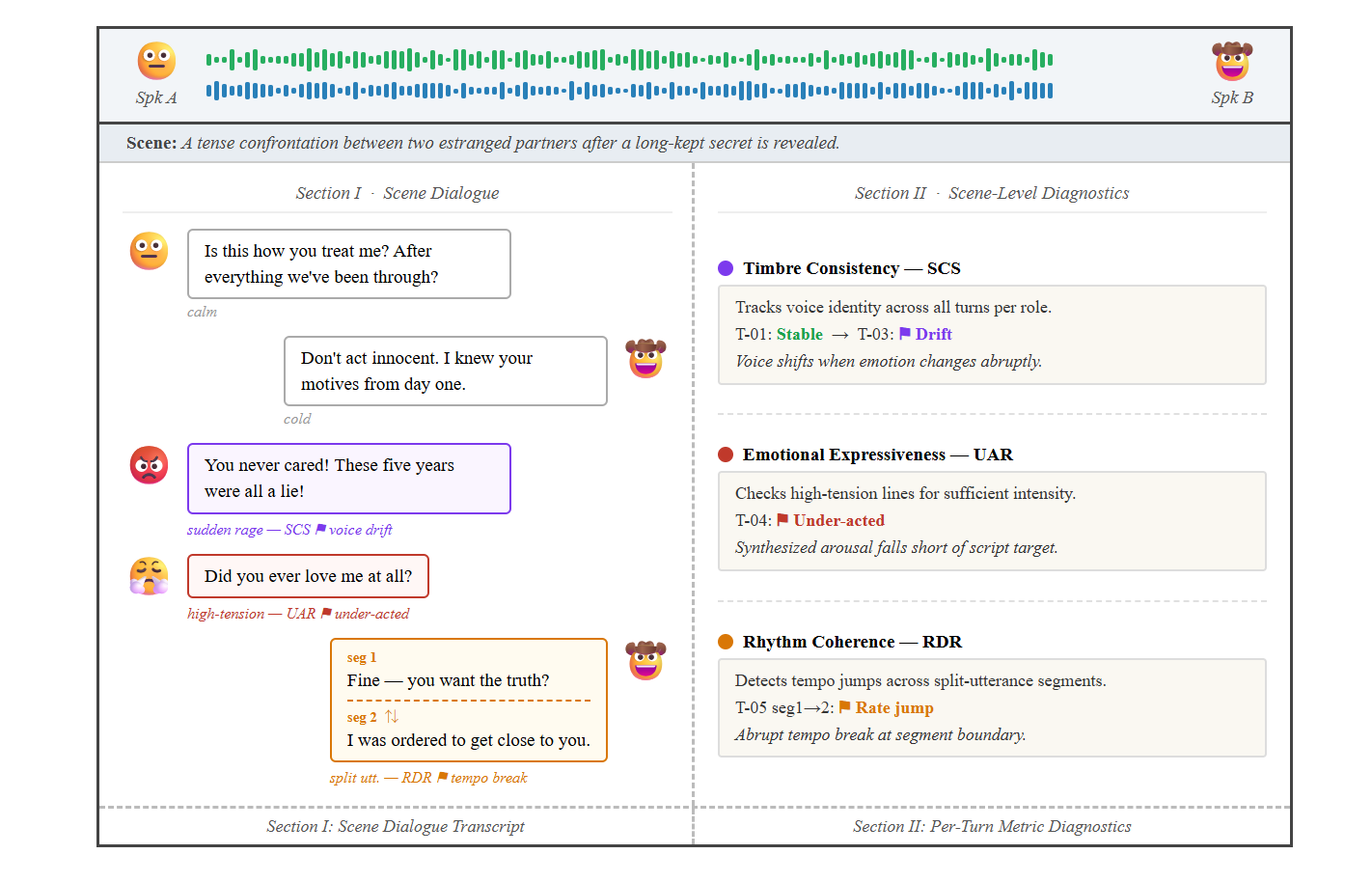}
  \caption{An example of SceneTTS-Bench evaluation showing scene-level diagnostics for a high-tension dialogue. The figure includes metrics for timbre consistency, emotional expressiveness, and rhythm coherence, as well as per-utterance diagnostics.}
  \label{fig:exp_data}
\end{figure*}

Existing TTS evaluation, however, takes the single utterance as its
atomic unit \cite{cao2025scores}. Metrics such as MOS, PESQ, and pairwise speaker similarity
assess each utterance in isolation \cite{jia2018transfer}; benchmarks like
LibriTTS~\cite{zen2019libritts}, VoiceMOS
Challenge~\cite{huang2022voicemos}, and SOMOS~\cite{maniati2022somos}
aggregate these scores into corpus-level statistics. This leaves three
failure modes critical to drama dubbing unexamined: no protocol checks
whether a character's timbre \emph{drifts} over a scene; emotional
assessment treats every utterance as equally important, missing
under-acting on high-tension utterances \cite{liu2024emotion}; and speech-rate evaluation
reports global means that mask local discontinuities between
adjacent synthesized segments \cite{sharma2021intra}.

These gaps cannot be closed by straightforward extensions of existing
metrics. Speaker similarity compares one utterance against one reference,
but detecting drift across dozens of turns demands population-level
distributional analysis \cite{dang2026tada}. Emotion classification tells us how an
utterance sounds, not whether it should have been more intense,
which requires structured tension supervision from the script \cite{cha2025jelly}.
Speech-rate means mask the abrupt jumps at segment boundaries that matter
most in spliced long-form synthesis \cite{hu2026qwen3}. Scene-level dubbing therefore calls
for purpose-built evaluation pipelines.

We present \textsc{SceneTTS-Bench}, a benchmark that reframes TTS
evaluation around the scene-level demands of drama dubbing
\cite{clark2019evaluating}. Rather than scoring utterances in isolation,
it evaluates whether a system can sustain character identity, deliver
sufficient emotional intensity, and preserve rhythmic continuity across a
scene. To enable fair comparison across systems with heterogeneous
conditioning interfaces, we introduce a backend-agnostic execution protocol
based on a Canonical Intermediate Representation (Canonical~IR), so that
different TTS backends operate under semantically equivalent inputs.
Building on this protocol, SceneTTS-Bench defines three automatic
scene-level metrics: Speaker Consistency Score~(SCS) for timbre drift,
Under-Acting Ratio~(UAR) for insufficient emotional expression against
script-level tension targets \cite{liang2025ece}, and Rate Discontinuity
Ratio~(RDR) for local speech-rate discontinuities across segmented
synthesis \cite{fujita2024speech}. Together, they expose concrete failure
modes rather than collapsing performance into one score. New backends can
be added through the common protocol without changing the evaluation.

Our contributions are summarized as follows:
\begin{itemize}
  \setlength{\itemsep}{1pt}
  \setlength{\parsep}{0pt}
  \setlength{\topsep}{3pt}
  \item A \textbf{scene-level evaluation paradigm} for drama dubbing TTS
        that shifts assessment from isolated utterances to multi-role,
        multi-turn dialogue, organized around timbre consistency,
        emotional expressiveness, and rhythm coherence---three capabilities critical to dubbing yet absent from existing benchmarks.

  \item A \textbf{backend-agnostic execution protocol} centered on a
        Canonical IR and dispatch adapter that enables fair,
        reproducible, and extensible comparison across TTS systems with
        heterogeneous conditioning interfaces.

  \item \textbf{Three automatic evaluation pipelines} producing
        interpretable per-utterance diagnostics for timbre drift (SCS),
        under-acting (UAR), and rate discontinuity (RDR).
\end{itemize}

\section{Related Work}

\subsection{TTS Benchmarks}
Existing TTS benchmarks mainly focus on sentence-level evaluation, emphasizing naturalness, intelligibility, MOS prediction, or speaker similarity, as in LibriTTS \cite{zen2019libritts}, VoiceMOS Challenge \cite{huang2022voicemos}, SOMOS \cite{maniati2022somos}, and Seed-TTS-Eval. While these benchmarks cover important utterance-level properties, their scoring unit remains the individual utterance, and speaker-related evaluation is typically formulated as pairwise similarity between a synthesized utterance and a reference. They therefore cannot directly assess stability across a role's dialogue turns; SCS measures this consistency.

\subsection{Emotion Evaluation}
Emotion evaluation in TTS is commonly conducted through categorical emotion recognition or MOS-style judgments of expressiveness and appropriateness, while speech emotion analysis often adopts dimensional representations such as arousal, valence, and dominance \cite{busso2025msp, martinez2020msp}. These approaches can determine what emotion is expressed, but they generally do not evaluate whether the synthesized speech reaches the dramatic intensity required by the script. Existing protocols particularly lack structured supervision for identifying under-acting on high-tension utterances. This motivates our Under-Acting Ratio (UAR), which combines dimensional emotion prediction with script-derived tension.

\subsection{Rhythm Evaluation in Long-Form TTS}
Prior work on long-form and context-aware TTS evaluation has shown that sentence-by-sentence assessment can miss important contextual effects \cite{omahony2021factors, guo2021conversational, clark2019evaluating}. At the same time, rhythm-related evaluation has mainly focused on sentence-internal prosody, duration control, or global speech-rate statistics \cite{gutierrez2021enhancing}. Such measures are useful for broad style characterization, but they are not designed to detect local rate discontinuities introduced when a long utterance is split into multiple synthesis calls. This motivates our Rate Discontinuity Ratio (RDR), which explicitly evaluates segment-boundary rhythm coherence within split long-form utterances.

\begin{figure*}[t]
  \centering
  \includegraphics[width=\textwidth]{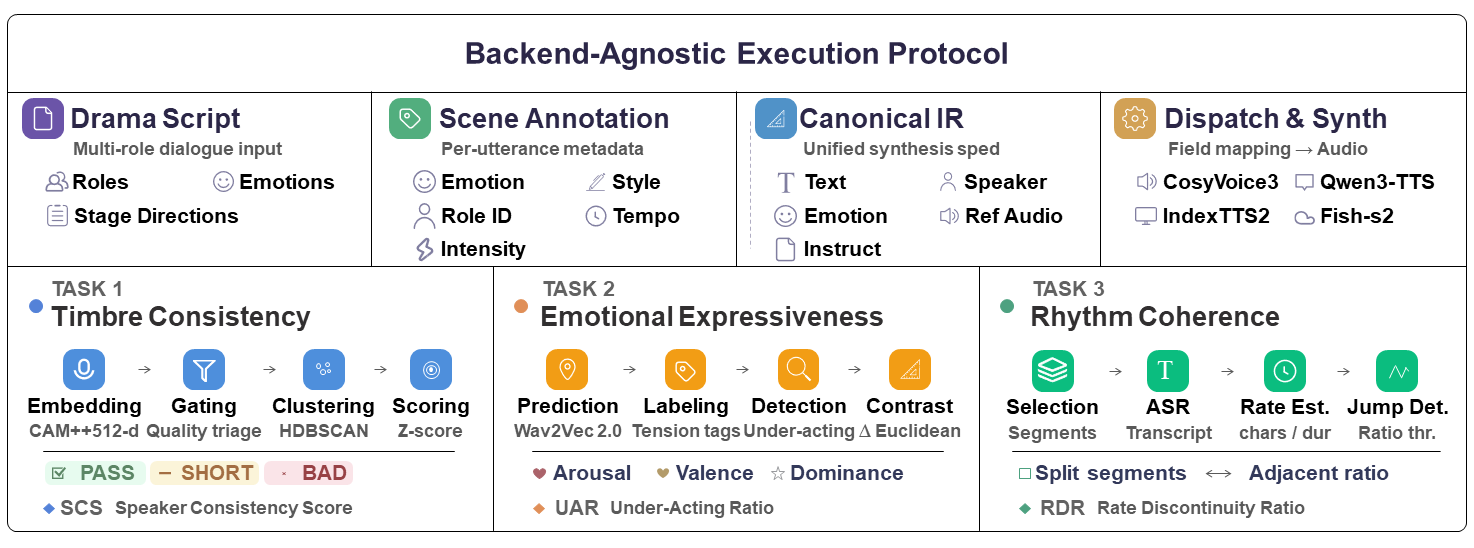}
  \caption{Overview of SceneTTS-Bench. \textbf{Top:} the backend-agnostic
  execution protocol converts drama scripts into a Canonical IR and
  dispatches to four TTS systems under identical conditions.
  \textbf{Bottom:} three evaluation pipelines assess timbre consistency
  (SCS), emotional expressiveness (UAR), and rhythm coherence (RDR),
  each producing per-utterance diagnostics.}
  \label{fig:overview}
\end{figure*}

\section{Method}
\label{sec:method}

\subsection{Overview}
\label{sec:overview}

SceneTTS-Bench is organized into an execution layer that standardizes
synthesis and an evaluation layer that measures the resulting audio
(Figure~\ref{fig:overview}) \cite{ren2026evaluating}. The execution layer converts drama scripts into a Canonical IR and dispatches to each system's native API, so that output differences reflect model capability rather than input discrepancy. The evaluation layer runs three automatic pipelines: Speaker Consistency Score (SCS) for timbre drift, Under-Acting Ratio (UAR) for under-acting on high-tension utterances, and Rate Discontinuity Ratio (RDR) for segment-boundary rate discontinuities \cite{liu2026prosodic}. We operationalize rhythm coherence as speech-rate continuity at segment boundaries. All pipelines produce per-utterance diagnostics alongside aggregated scores.

\subsection{Backend-Agnostic Execution Protocol}
\label{sec:protocol}

Different TTS systems condition on different inputs (reference audio,
natural-language instructions, or categorical style tags) \cite{qiang2025instructaudio}. Invoking each
through its native interface with independently prepared inputs would
conflate model capability with input discrepancy. We therefore define a
Canonical Intermediate Representation (Canonical IR): a fixed tuple of
five fields (text, speaker, emotion, reference audio, natural-language
instruction) chosen as the minimal superset covering all included
systems. Scene annotation first extracts per-utterance metadata from
dialogue context and stage directions; a lightweight dispatch adapter then
maps each Canonical IR entry to the target system's API, so that
CosyVoice3\cite{du2025cosyvoice}, Qwen3-TTS\cite{hu2026qwen3}, IndexTTS2\cite{zhou2026indextts2}, and Fish-S2\cite{liao2026fishaudios2} all operate on the same
semantic specification.

The protocol defines a common task rather than a separate optimized recipe
for each backend. An adapter may translate or omit unsupported fields, but
cannot introduce semantic information unavailable to other systems. Thus,
results measure how each backend realizes the same role, emotion, and timing
specification, rather than the best performance attainable through bespoke
prompt engineering. A new backend requires only a dispatch mapping; the
corpus and downstream metrics remain unchanged.

\subsection{Timbre Consistency}
\label{sec:timbre}

In drama dubbing, a character may speak across dozens of turns, and
listeners expect the voice to remain stable. TTS systems can exhibit
gradual timbre drift or sudden identity collapse as dialogue progresses.
Conventional pairwise speaker similarity compares one utterance against
one reference and cannot reveal distributional shifts across a full
role population. Our pipeline operates at the role level: we extract a
speaker embedding for every utterance of a role using CAM++
\cite{wang23ha_interspeech}, apply quality gating to remove utterances
shorter than 0.5\,s or with degraded audio quality (e.g., silence or
excessive noise), and cluster the remaining embeddings with HDBSCAN
\cite{campello2013density} to identify the dominant voice cluster with
centroid $\boldsymbol{\mu}_r$. If the largest cluster is too small under
cross-script grouping, a robust core is constructed by selecting the
samples closest to the global centroid and rebuilding the dominant centroid
from this core set.

Each utterance's cosine similarity to the dominant centroid is computed as
$s_i = \cos(\mathbf{e}_i, \boldsymbol{\mu}_r)$. Let $\mu_s$ and $\sigma_s$
denote the mean and standard deviation of $\{s_i\}$ within the dominant
cluster. A z-score based penalty is applied to each utterance:
\begin{equation}
  Z_i = \frac{\mu_s - s_i}{\sigma_s}
  \label{eq:zscore}
\end{equation}
Only utterances deviating beyond a tolerance band incur a loss:
\begin{equation}
  L_i = \max(0,\; Z_i - \kappa)
  \label{eq:loss}
\end{equation}
where $\kappa = 2$ is the tolerance threshold (penalties begin only when
a sample falls more than $2\sigma_s$ below the cluster mean). The
role-level Speaker Consistency Score is then:
\begin{equation}
  \mathrm{SCS} = \exp\!\Bigl(
    -\,\alpha \cdot \frac{1}{N}\sum_{i=1}^{N} L_i
  \Bigr)
  \label{eq:scs}
\end{equation}
where $N$ is the number of quality-gated utterances and $\alpha$ is a
penalty strength coefficient (default $\alpha = 5$). SCS lies in $(0,1]$:
values near 1 indicate that all utterances cluster tightly around the
dominant voice, while low values signal substantial timbre drift.

This construction penalizes departures from the voice a system most
consistently realizes, without assuming that one reference represents every
speaking condition. Quality gating and the $2\sigma_s$ tolerance band reduce
false alarms from silence, noise, and ordinary expressive variation. The
utterance losses remain available so a low SCS can be traced to specific
dialogue turns.

\subsection{Emotional Expressiveness}
\label{sec:emotion}

For drama dubbing, the key question is not whether synthesized speech
conveys \emph{some} emotion, but whether it reaches the intensity
required by the script on high-tension utterances. We therefore combine
script-side supervision with prediction-side emotion estimation. Each
utterance is assigned a tension label
(\texttt{high\_arousal}, \texttt{high\_dominance}, or \texttt{normal}),
and only the first two participate in under-acting evaluation. We use a
Wav2Vec~2.0 dimensional emotion model \cite{wagner2023dawn} to predict
arousal $\hat{a}_u$ and dominance $\hat{d}_u$ for every utterance.

Under-acting is defined by a threshold check on the axis specified by the
tension label:
\begin{equation}
  \mathrm{underact}(u) =
  \begin{cases}
    \mathbb{1}[\,\hat{a}_u < \tau_a\,], & \text{if label} = \texttt{high\_arousal} \\
    \mathbb{1}[\,\hat{d}_u < \tau_d\,], & \text{if label} = \texttt{high\_dominance}
  \end{cases}
  \label{eq:underact}
\end{equation}
where $\tau_a$ and $\tau_d$ are arousal and dominance thresholds. The
aggregated Under-Acting Ratio is
\begin{equation}
  \mathrm{UAR} = \frac{
    \sum_{u \in \mathcal{H}_a} \mathbb{1}[\hat{a}_u < \tau_a]
    \;+\;
    \sum_{u \in \mathcal{H}_d} \mathbb{1}[\hat{d}_u < \tau_d]
  }{
    |\mathcal{H}_a| + |\mathcal{H}_d|
  }
  \label{eq:uar}
\end{equation}
where $\mathcal{H}_a$ and $\mathcal{H}_d$ denote utterances labeled
\texttt{high\_arousal} and \texttt{high\_dominance}, respectively. UAR
thus measures how often a system falls below the script-required
emotional intensity. Scenes without high-tension utterances are excluded
from UAR aggregation.

UAR is intentionally one-sided: it detects failure to reach required
intensity, but neither rewards exaggeration nor scores normal-tension lines.
It is therefore an under-acting diagnostic, not a complete measure of
emotional naturalness. Restricting the denominator to high-tension lines
prevents abundant neutral dialogue from concealing consequential failures.

\subsection{Rhythm Coherence}
\label{sec:rhythm}

In drama dubbing, long utterances are often synthesized in multiple
segments, which can introduce local rate discontinuities even when each
segment sounds natural in isolation \cite{eskimez2024total, galdino2025impact}. We therefore evaluate rhythm coherence at the segment-group level by linking all segments originating from the same source utterance.

For each segment $i$, an ASR module provides token count $n_i$, and VAD
yields the active speech duration $d_i^{\mathrm{act}}$, excluding leading
and trailing silence. We define the segment speech rate as
\begin{equation}
  s_i = \frac{n_i}{d_i^{\mathrm{act}}}
  \label{eq:rate}
\end{equation}
where $n_i$ is character count for Chinese and word count for English
\cite{yang2025evaluating}. Segments with insufficient token count or active duration are discarded.

Within each valid group, we compare all segment pairs and flag a rate
jump when their rate ratio exceeds a threshold:
\begin{equation}
  \mathrm{jump}(i, j) =
    \mathbb{1}\!\left[\,
      \max\!\left(\frac{s_i}{s_j},\;\frac{s_j}{s_i}\right)
      > \tau_r
    \,\right]
  \label{eq:jump}
\end{equation}

The Rate Discontinuity Ratio is then defined as the fraction of valid
groups containing at least one flagged pair:
\begin{equation}
  \mathrm{RDR} = \frac{|\{g \in \mathcal{G} : \exists\,(i,j)\in g,\;
    \mathrm{jump}(i,j)=1\}|}{|\mathcal{G}|}
  \label{eq:rdr}
\end{equation}
where $\mathcal{G}$ denotes the set of valid segment groups. We aggregate
at the group level because any single jump can make the reconstructed
utterance perceptually discontinuous. Scenes without valid groups are
excluded from RDR aggregation.

Pairwise comparison preserves local discontinuities that could cancel in an
average, while group-level aggregation prevents heavily segmented utterances
from dominating the metric. RDR is therefore the fraction of reconstructed
utterances containing a suspicious rate transition; stored pairwise ratios
identify the boundary for inspection. It complements broader prosody
measures because pitch, energy, pauses, and coarticulation are not modeled.

\section{Experiments}

\subsection{Experimental Setup}
\label{sec:setup}

The benchmark contains two subsets: a primary \emph{real-world} subset of
50 Chinese and 50 English scenes from publicly available scripts (IMSDb
under fair use; Project Gutenberg Drama in the public domain)
\cite{ramakrishna2017linguistic}, and a supplementary \emph{generated}
subset of 30 Chinese and 30 English scenes produced programmatically
across ten genres \cite{wang2024rolellm}. In total, the corpus comprises
160 scenes and approximately 10{,}300 utterances. As shown in
Table~\ref{tab:data-stats}, the two subsets differ substantially in
structure: real-world Chinese scenes have more roles, real-world English
has fewer high-tension lines, and split-utterance prevalence ranges from
7.9\% to 32.0\%. These contrasts test generalization beyond one scene
configuration. Scripts, annotations, and licensed outputs are released
under CC-BY-4.0.

GPT-5.2 annotates role, emotion, tension, style, and tempo
\cite{gilardi2023chatgpt}. Three trained annotators independently rate 300
stratified high-tension lines; a separate 500-line audit obtains 99.6\%
agreement for role identity and 91.2\% for emotion labels.

We evaluate four TTS systems: CosyVoice3 \cite{du2025cosyvoice},
Qwen3-TTS \cite{hu2026qwen3}, IndexTTS2 \cite{zhou2026indextts2}, and
Fish-S2 \cite{liao2026fishaudios2}. Each synthesizes the full corpus via
the Canonical IR protocol (Section~\ref{sec:protocol}), producing about
41{,}300 utterances. Speaker embeddings are extracted
with CAM++, dimensional emotion is predicted by a Wav2Vec~2.0 regressor
trained on MSP-Podcast \cite{lotfian2017building}, and thresholds are set
to $\tau_a = 0.75$, $\tau_d = 0.75$, and $\tau_r = 1.5$. Quality gating
removes utterances shorter than 0.5\,s or degraded audio
\cite{reddy2021dnsmos}. Thirty gender-balanced reference timbres (15 per
language) serve as voice anchors.

\begin{table}[t]
  \caption{Corpus statistics by subset and language. Values are per-scene
  means with standard deviations in parentheses.}
  \label{tab:data-stats}
  \centering
  \small
  \setlength{\tabcolsep}{2.5pt}
  \begin{tabularx}{\columnwidth}{@{}>{\raggedright\arraybackslash}p{0.25\columnwidth}YYYY@{}}
    \toprule
    & \textbf{Real-ZH} & \textbf{Real-EN}
    & \textbf{Gen-ZH}  & \textbf{Gen-EN} \\
    \midrule
    Scenes          & 50 & 50 & 30 & 30 \\
    Roles / scene   & 9.3\,(1.6) & 3.2\,(0.7) & 3.0\,(0.0) & 3.0\,(0.0) \\
    Turns / scene   & 59.9\,(0.4) & 58.0\,(3.1) & 77.0\,(11.7) & 70.4\,(8.8) \\
    Utt.\ len (s)   & 5.7\,(4.8) & 3.8\,(3.5) & 4.1\,(2.0) & 3.6\,(1.7) \\
    High-tension \% & 14.4\,(4.3)\% & 4.1\,(2.5)\% & 21.3\,(2.7)\% & 20.5\,(1.9)\% \\
    Split-seg.\ \%  & 7.9\,(4.6)\% & 16.8\,(14.3)\% & 25.3\,(6.4)\% & 32.0\,(10.4)\% \\
    \bottomrule
  \end{tabularx}
\end{table}

\subsection{Overall Comparison}
\label{sec:overall}

Table~\ref{tab:main} presents the main results across the three
evaluation dimensions. A clear pattern is that no single system
dominates all three. Qwen3-TTS achieves the highest timbre consistency
(SCS\,=\,0.96) and the best rhythm coherence (RDR\,=\,17.1\%), yet
shows the worst emotional expressiveness (UAR\,=\,85.5\%). By contrast,
IndexTTS2 achieves the lowest under-acting ratio (UAR\,=\,81.1\%) but
only third-best timbre consistency (SCS\,=\,0.74), while Fish-S2 ranks
last on both SCS and RDR. CosyVoice3 remains competitive across all
dimensions without a pronounced weakness. These cross-metric trade-offs
show that scene-level evaluation exposes capability differences that
cannot be summarized by a single ranking \cite{tan2021survey}. Most
pairwise differences in Table~\ref{tab:main} are statistically
significant ($p<0.05$, 1{,}000 scene-level bootstrap resamples)
\cite{berg2012empirical}, with only two exceptions: CosyVoice3 vs.\
IndexTTS2 on UAR (0.5\,pp, $p=0.12$) and IndexTTS2 vs.\ Fish-S2 on SCS
(0.03, $p=0.09$).

Keeping the dimensions separate is operationally useful: a production team
can distinguish identity failures from under-expression or segmentation
artifacts and choose a backend according to the dominant risk in its material.

\begin{table}[t]
  \caption{Main results with 95\% bootstrap confidence intervals.
  Best results are in \textbf{bold}.}
  \label{tab:main}
  \centering
  \small
  \setlength{\tabcolsep}{3pt}
  \begin{tabularx}{\columnwidth}{@{}>{\raggedright\arraybackslash}XYYY@{}}
    \toprule
    \textbf{System}
    & SCS$\uparrow$ & UAR(\%)$\downarrow$ & RDR(\%)$\downarrow$ \\
    \midrule
    CosyVoice3
    & $.90{\scriptstyle\pm.02}$
    & $81.6{\scriptstyle\pm2.3}$
    & $27.6{\scriptstyle\pm2.4}$ \\
    Qwen3-TTS
    & $\mathbf{.96}{\scriptstyle\pm.01}$
    & $85.5{\scriptstyle\pm2.1}$
    & $\mathbf{17.1}{\scriptstyle\pm1.8}$ \\
    IndexTTS2
    & $.74{\scriptstyle\pm.03}$
    & $\mathbf{81.1}{\scriptstyle\pm2.4}$
    & $21.4{\scriptstyle\pm2.0}$ \\
    Fish-S2
    & $.71{\scriptstyle\pm.03}$
    & $84.5{\scriptstyle\pm2.6}$
    & $32.1{\scriptstyle\pm2.5}$ \\
    \bottomrule
  \end{tabularx}
\end{table}

\textbf{Consistency across real-world and generated scripts.}
Table~\ref{tab:real-vs-gen} breaks down results by script source.
Despite the substantial structural differences in
Table~\ref{tab:data-stats}, system rankings are preserved across both
subsets on SCS and RDR; on UAR, CosyVoice3 and IndexTTS2 swap adjacent
ranks in the generated subset (79.8 vs.\ 79.9), a negligible difference
consistent with their overlapping confidence intervals in
Table~\ref{tab:main}. Overall, this broad agreement supports
generalization beyond any single corpus configuration.

Absolute scores vary in expected ways. SCS is lower on real-world
scripts, where scenes contain more roles and thus impose a harder
timbre-maintenance setting than the fixed three-role template of the
generated subset. UAR is generally higher on real-world scripts, though
the real-world English split contains relatively few high-tension
utterances (4.1\%) and therefore provides a smaller evaluation sample.
RDR varies moderately with the proportion of split utterances, with the
generated English subset yielding the highest values. Even so, the
system ordering remains stable, suggesting that the main differences are
driven by backend behavior rather than corpus composition. Uniformly high
UAR values (81--86\%) also reveal that high-tension dialogue remains
substantially under-expressed.

\begin{table}[t]
  \caption{Scene-level metrics by script source. R: real-world scripts;
  G: generated scripts.}
  \label{tab:real-vs-gen}
  \centering
  \small
  \setlength{\tabcolsep}{2pt}
  \begin{tabularx}{\columnwidth}{@{}>{\raggedright\arraybackslash}p{0.23\columnwidth}YYYYYY@{}}
    \toprule
    & \multicolumn{2}{c}{\textbf{SCS}$\uparrow$}
    & \multicolumn{2}{c}{\textbf{UAR}$\downarrow$}
    & \multicolumn{2}{c}{\textbf{RDR}$\downarrow$} \\
    \cmidrule(lr){2-3} \cmidrule(lr){4-5} \cmidrule(lr){6-7}
    \textbf{System} & R & G & R & G & R & G \\
    \midrule
    CosyVoice3 & 0.88 & 0.93 & 82.7 & 79.8 & 28.5  & 26.1 \\
    Qwen3-TTS  & 0.95 & 0.98 & 86.4 & 84.0 & 17.8 & 15.9 \\
    IndexTTS2  & 0.72 & 0.78 & 81.8 & 79.9 & 22.1 & 20.2 \\
    Fish-S2    & 0.68 & 0.76 & 85.4 & 83.0 & 33.5 & 29.8 \\
    \bottomrule
  \end{tabularx}
\end{table}

\subsection{Scene-Level vs.\ Sentence-Level Metrics}
\label{sec:scene-vs-sent}

\begin{figure}[t]
  \centering
  \includegraphics[width=0.95\columnwidth]{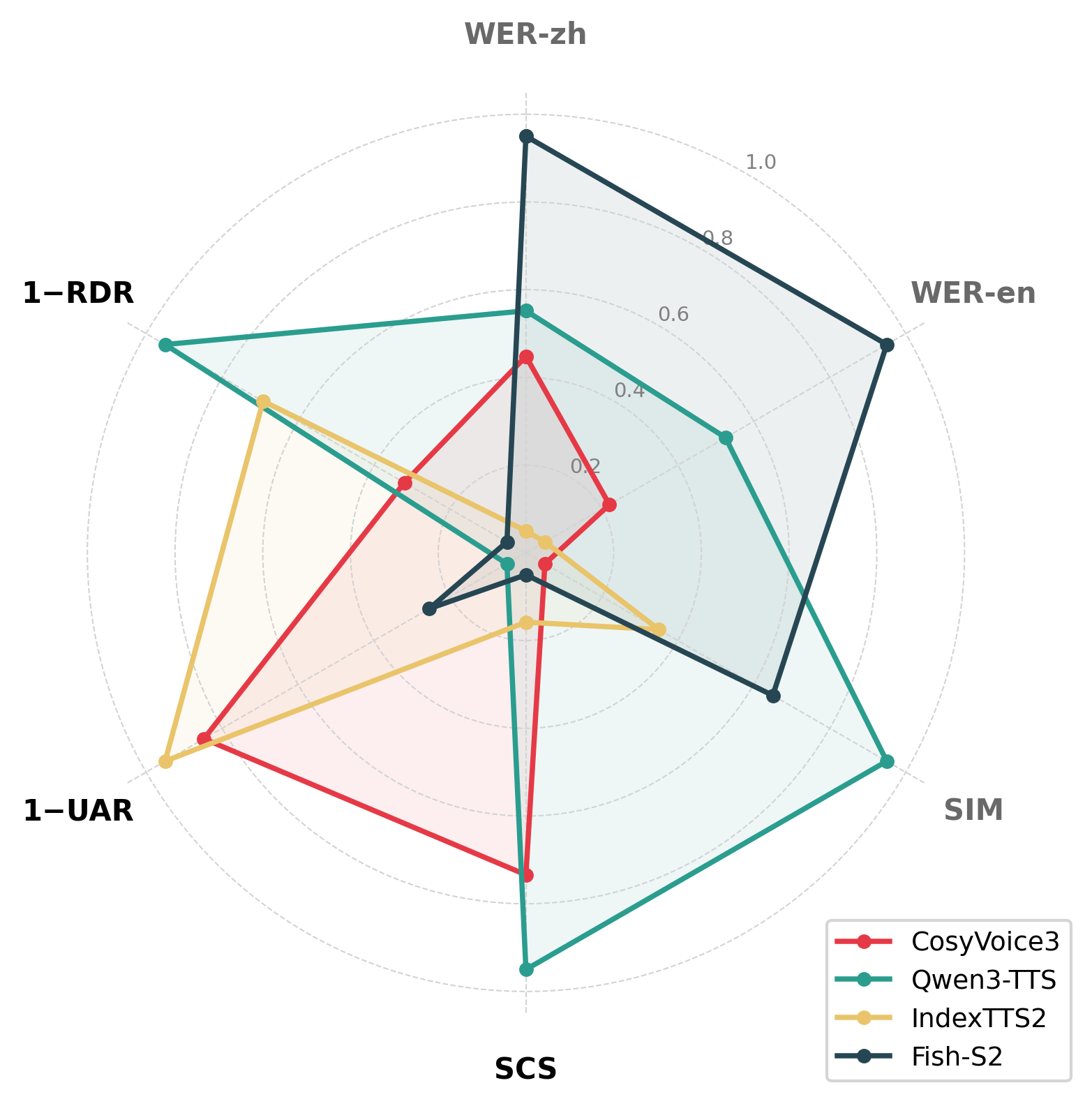}
  \caption{Normalized sentence-level (grey labels) and scene-level
  (black labels) profiles; higher is better on every axis.}
  \label{fig:radar}
\end{figure}

We compare system rankings from three sentence-level metrics
(WER-zh, WER-en, SIM following Seed-TTS-Eval~\cite{anastassiou2024seed})
with our scene-level pipelines (Table~\ref{tab:scene-vs-sent},
Figure~\ref{fig:radar}).

Rankings diverge substantially. Fish-S2 ranks 1st on both WER measures
yet 4th on SCS and RDR, a dramatic inversion invisible to sentence-level
evaluation alone. Qwen3-TTS achieves the highest SIM (rank~1) and leads
both SCS (rank~1) and RDR (rank~1), yet ranks worst on UAR (rank~4),
showing that strong reference matching and rhythmic smoothness do not
guarantee emotional intensity \cite{li2023styletts}. IndexTTS2 ranks last
on WER (rank~4) but best on UAR (rank~1), revealing that intelligibility
and expressiveness are largely independent dimensions. CosyVoice3 ranks
mid-pack on all metrics without a pronounced extreme.
These inversions expose failure modes inaccessible to conventional
sentence-level metrics \cite{maiti2023speechlmscore}. Human correlations
($\rho=0.61$--$0.78$; Section~\ref{sec:human-corr}) further confirm their
perceptual relevance to dubbing.

\begin{table}[t]
  \caption{Sentence- and scene-level metric ranks.}
  \label{tab:scene-vs-sent}
  \centering
  \small
  \setlength{\tabcolsep}{1.5pt}
  \begin{tabularx}{\columnwidth}{@{}>{\raggedright\arraybackslash}p{0.18\columnwidth}YYYYYY@{}}
    \toprule
    & \multicolumn{3}{c}{\textbf{Sentence-Level}}
    & \multicolumn{3}{c}{\textbf{Scene-Level}} \\
    \cmidrule(lr){2-4} \cmidrule(lr){5-7}
    \textbf{System}
    & WER-zh$\downarrow$ & WER-en$\downarrow$ & SIM$\uparrow$
    & SCS$\uparrow$ & UAR$\downarrow$ & RDR$\downarrow$ \\
    \midrule
    CosyVoice3
    & 0.83$_3$ & 1.42$_3$ & 0.761$_4$
    & 0.90$_2$ & 81.6$_2$ & $27.6_3$ \\
    Qwen3-TTS
    & 0.77$_2$ & 1.24$_2$ & \textbf{0.794}$_1$
    & $\mathbf{0.96}_1$ & 85.5$_4$ & $\mathbf{17.1}_1$ \\
    IndexTTS2
    & 1.06$_4$ & 1.52$_4$ & 0.772$_3$
    & 0.74$_3$ & \textbf{81.1}$_1$ & $21.4_2$ \\
    Fish-S2
    & \textbf{0.54}$_1$ & \textbf{0.99}$_1$ & 0.783$_2$
    & 0.71$_4$ & 84.5$_3$ & $32.1_4$ \\
    \bottomrule
  \end{tabularx}
\end{table}

\subsection{Correlation with Human Judgments}
\label{sec:human-corr}

We assess alignment with human perception through a controlled study.

\textbf{Protocol.}
We draw 800 units (200 per system), stratified by tension, segment type,
and language. Timbre uses 5--8-turn role trajectories, rhythm uses adjacent
segments, and emotion uses utterances with dialogue context. Five
experienced annotators rate each unit on five-point Likert scales;
Krippendorff's $\alpha$ is 0.71, 0.65, and 0.68 for timbre, emotion, and
rhythm, respectively.

\textbf{Results.}
Table~\ref{tab:human-corr} reports per-utterance and system-level
correlations. SCS aligns most strongly ($\rho=0.782$), followed by RDR
($\rho=0.724$), whose rate-ratio threshold reflects perceived
discontinuities. UAR shows moderate correlation ($\rho=0.613$), consistent
with the subjectivity of emotional intensity \cite{scherer2003vocal};
system-level rankings are nevertheless preserved. All three metrics are
useful at utterance and system granularity.

Agreement at both utterance and system levels serves different uses: local
scores support diagnosis, while aggregated scores expose stable backend
tendencies for comparative benchmarking.

\begin{table}[t]
  \caption{Correlation with human ratings over 800 samples
  ($p<0.001$); system-level $\rho$ uses four backend means.}
  \label{tab:human-corr}
  \centering
  \small
  \setlength{\tabcolsep}{2pt}
  \begin{tabularx}{\columnwidth}{@{}>{\raggedright\arraybackslash}p{0.18\columnwidth}>{\raggedright\arraybackslash}p{0.15\columnwidth}YYYY@{}}
    \toprule
    & & \multicolumn{3}{c}{\textbf{Utterance-Level}}
    & \textbf{Sys.} \\
    \cmidrule(lr){3-5} \cmidrule(lr){6-6}
    \textbf{Dimension} & \textbf{Metric}
    & Spear. & Pear. & Kend.
    & $\rho$ \\
    \midrule
    Timbre    & SCS
    & 0.782 & 0.768 & 0.604
    & 1.00 \\
    Emotion   & UAR
    & 0.613 & 0.587 & 0.451
    & 1.00 \\
    Rhythm    & RDR
    & 0.724 & 0.695 & 0.547
    & 0.80 \\
    \bottomrule
  \end{tabularx}
\end{table}

\subsection{Scope, Limitations, and Intended Use}
\label{sec:limitations}

SceneTTS-Bench diagnoses three recurrent failures rather than defining
dubbing quality completely. SCS may respond to recording conditions or
vocal effort; UAR does not judge naturalness or over-acting; RDR omits pitch,
energy, pause, and coarticulation discontinuities. The correlations in
Table~\ref{tab:human-corr} support these proxies, but listening tests remain
necessary for final production decisions.

Generated scenes supplement rather than replace real dialogue, and the
tested voices do not cover every accent, age group, style, or recording
condition. New languages and domains require local validation of
annotations, predictors, embeddings, and thresholds. We recommend reporting
SCS, UAR, and RDR separately with confidence intervals and per-utterance
diagnostics: a single score would hide the trade-offs in
Table~\ref{tab:main}. Human review remains the authority for artistic quality
and deployment suitability.

The benchmark supports model comparison and failure analysis, not automatic
acceptance of released audio. Because its thresholds are corpus-calibrated
operating points, transfers to new languages or genres require a locally
rated validation set; authors should also document sensitivity to calibration.

\section{Conclusion}
\label{sec:conclusion}

We presented SceneTTS-Bench, a benchmark that shifts TTS evaluation for
drama dubbing from the sentence level to the scene level. Built on a
backend-agnostic Canonical IR, it evaluates timbre consistency, emotional
expressiveness, and rhythm coherence through three automatic pipelines with
per-utterance diagnostics. A bilingual corpus of 160 scenes, together with
validated LLM-generated tension labels, supports reliable evaluation in
this setting. Experiments on four systems show that no single backend
dominates all three dimensions and that scene-level rankings differ
substantially from sentence-level metrics. This suggests that evaluating
dubbing-oriented TTS requires more than isolated utterance quality: it
requires measuring whether a system can sustain character identity,
dramatic intensity, and temporal continuity across a scene. More broadly,
this shift elevates scene-level capability from an implicit by-product of
synthesis quality to an explicit target of evaluation and future system
design. We hope SceneTTS-Bench can serve as a practical foundation for more
holistic, diagnostic, and application-aligned evaluation of synthesized
speech.

\clearpage
\balance
\bibliography{references}
\bibliographystyle{ACM-Reference-Format}

\end{document}